\documentclass[12pt]{article}
\usepackage{graphicx,amsfonts,amsmath,amssymb,amsthm,mathrsfs,url,hyperref}
\usepackage[usenames,dvipsnames]{xcolor}

\title{Complex Charges, Time Reversal Asymmetry,\\ and Interior-Boundary Conditions in\\ Quantum Field Theory}
\author{
Julian Schmidt\footnote{Mathematisches Institut,
     Eberhard-Karls-Universit\"at, Auf der Morgenstelle 10, 72076
     T\"ubingen, Germany}\ \footnote{E-mail:
     juls@maphy.uni-tuebingen.de}\ \  and
Roderich Tumulka$^*$\footnote{E-mail:
     roderich.tumulka@uni-tuebingen.de}
}
\date{October 4, 2018}

\newcommand{\Hilbert}{\mathscr{H}}

\newcommand{\Q}{\mathcal{Q}}
\renewcommand{\Re}{\mathrm{Re}}
\renewcommand{\Im}{\mathrm{Im}}
\newcommand{\RRR}{\mathbb{R}}
\newcommand{\CCC}{\mathbb{C}}
\newcommand{\SSS}{\mathbb{S}}
\newcommand{\ZZZ}{\mathbb{Z}}

\newcommand{\free}{{\mathrm{free}}}
\newcommand{\rev}{{\mathrm{rev}}}
\newcommand{\ve}{\boldsymbol{e}}
\newcommand{\vj}{\boldsymbol{j}}

\newcommand{\vx}{\boldsymbol{x}}
\newcommand{\vy}{\boldsymbol{y}}

\newcommand{\vY}{\boldsymbol{Y}}

\newcommand{\vomega}{\boldsymbol{\omega}}

\newcommand{\be}{\begin{equation}}
\newcommand{\ee}{\end{equation}}

\begin{document}
\maketitle
\begin{abstract}
While fundamental physically realistic Hamiltonians should be invariant under time reversal, time \emph{asymmetric} Hamiltonians can occur as mathematical possibilities or effective Hamiltonians. Here, we study conditions under which non-relativistic Hamiltonians involving particle creation and annihilation, as come up in quantum field theory (QFT), are time asymmetric. It turns out that the time reversal operator $T$ can be more complicated than just complex conjugation, which leads to the question which criteria determine the correct action of time reversal. We use Bohmian trajectories for this purpose and show that time reversal symmetry can be broken when charges are permitted to be complex numbers, where ``charge'' means the coupling constant in a QFT that governs the strength with which a fermion emits and absorbs bosons. We pay particular attention to the technique for defining Hamiltonians with particle creation based on interior-boundary conditions, and we find them to generically be time asymmetric. Specifically, we show that time asymmetry for complex charges occurs whenever not all charges have equal or opposite phase. We further show that, in this case, the corresponding ground states can have non-zero probability currents, and we determine the effective potential between fermions of complex charge.

\medskip

  \noindent 
%
%
\end{abstract}

\section{Introduction}

It is well known that macroscopic irreversibility (i.e., the thermodynamic arrow of time) is due to special initial conditions while the fundamental laws of physics are time symmetric (i.e., symmetric against time reversal); see, e.g., \cite{Pen}.
In this paper, however, we also consider time \emph{asymmetric} Hamiltonians. We explore how non-relativistic Hamiltonians involving particle creation and annihilation can be or fail to be time symmetric. This question comes up naturally when studying Hamiltonians implemented using the technique of interior-boundary conditions (IBCs), as generic coefficients in an IBC will lead to time asymmetry. However, the issue is not limited to IBC Hamiltonians but can arise, as we show, in any situation with particle creation provided that the charges, i.e., the coupling constants in front of creation and annihilation terms in the Hamiltonian, are allowed to be complex rather than real. 

We also make the observation that the time reversal operator $T$ in non-relativistic quantum theories with particle creation is not necessarily given by complex conjugation, but more generally by complex conjugation followed by multiplication with a phase factor depending on the particle number. We use Bohmian trajectories to identify the right representation of time reversal, which is still an anti-unitary operator $T$, as one would expect from general and abstract considerations \cite{Uhl63,Rob}. This Bohmian approach then makes it possible to prove or disprove the time symmetry of a Hamiltonian under consideration. Specifically, we show of a Hamiltonian involving the emission and absorption of bosons by fermions with complex charges that it is time symmetric if and only if all charges have equal or opposite phases.

Time \emph{asymmetric} Hamiltonians are rather unfamiliar. For example, usual Hamiltonians of non-relativistic quantum mechanics without particle creation, i.e.,
\be\label{DeltaV}
H = -\Delta + V
\ee
with time-independent real-valued potential $V$ are always time symmetric. 
As we will discuss in Section~\ref{sec:BC}, also Hamiltonians with point interactions (Dirac delta potentials) and those defined by \eqref{DeltaV} using ordinary local boundary conditions such as Dirichlet and Neumann boundary conditions are always time symmetric. 

Cases of time asymmetric Hamiltonians are known to occur in the presence of an external magnetic field, as magnetic fields change sign under time reversal; the situation is parallel to the fact that the space translation invariance of the Hamiltonian is broken by any non-uniform external field, so that one would perhaps not regard the situation as a serious violation of time symmetry, as one may have in mind that external fields should be appropriately transformed as well. Likewise, when $V$ in \eqref{DeltaV} is time-dependent, then technically speaking time reversal invariance  is broken as $(\psi^*)_{-t} \neq (\psi_t)^*$, but it gets restored if we also transform $V(t)\mapsto V(-t)$, as we should for a $V$ including external fields. Another known case of a time asymmetric Hamiltonian \cite{BM} involves the Dirac equation in 2d with matrix-valued potential (a case with mathematical parallels to magnetic fields).

The QFTs we consider here are variants of the following basic scheme: $N$ fermions are fixed at the (pairwise distinct) locations $\vx_1,\ldots,\vx_N\in\RRR^3$ and emit and absorb non-relativistic, spinless bosons of mass $m>0$. That is, the Hilbert space is the bosonic Fock space over $L^2(\RRR^3)$,
\be\label{Hilbertdef}
\Hilbert = \bigoplus_{n=0}^\infty \mathrm{Sym}(L^2(\RRR^3)^{\otimes n})\,,
\ee
and wave functions $\psi\in\Hilbert$ can be regarded as complex-valued functions on the configuration space $\Q$ of a variable number of particles,
\be\label{Qdef}
\Q= \bigcup_{n=0}^\infty (\RRR^3)^n \,,
\ee
such that each sector of $\psi$ is invariant under permutations of the particles.
The Hamiltonian is, formally,
\be\label{Hdef}
H = H_\free + \sum_{j=1}^N \bigl( g_j \, a(\vx_j) + g_j^* \, a^\dagger(\vx_j) \bigr) \,. 
\ee  
Here, $H_\free$ is the second quantization of the 1-particle operator $-\frac{1}{2m}\Delta + E_0$ where $E_0\geq 0$ is the energy that must be expended for creating a boson; we set $\hbar=1$; $g_j\in \CCC\setminus \{0\}$ is a coupling constant that we will call the charge of fermion number $j$, and $a(\vx)$ and $a^\dagger(\vx)$ are the annihilation and creation operators at the location $\vx\in\RRR^3$,
\begin{align}
\bigl(a(\vx)\,\psi\bigr)(\vy_1...\vy_n) 
&= \sqrt{n+1}\; \psi\bigl(\vy_1...\vy_n,\vx\bigr)\label{adef1}\\
\bigl(a^\dagger(\vx)\,\psi\bigr)(\vy_1...\vy_n) 
&= \frac{1}{\sqrt{n}} \sum_{k=1}^n \delta^3(\vy_k-\vx)\,\psi\bigl(y\setminus \vy_k\bigr)\label{adef2}
\end{align}
with the notation $y\setminus \vy_k$ meaning $(\vy_1...\vy_{k-1},\vy_{k+1}...\vy_n)$ leaving out $\vy_k$.

One of our results is that this Hamiltonian is time symmetric if and only if any two $g_j, g_i$ have either equal or opposite phase, i.e.,
\be\label{phase}
\text{time symmetry} ~~~ \Leftrightarrow ~~~ 
g_i^* g_j \in \RRR\text{ for all }i,j=1...N\,.
\ee

Since the Hamiltonian \eqref{Hdef} is ultraviolet (UV) divergent, one may want to introduce an UV cutoff by smearing out the fermions using a square-integrable approximation $\varphi:\RRR^3\to\RRR$ to the Dirac delta function, which amounts to replacing the $a$ and $a^\dagger$ operators by
\begin{align}
\bigl(a_\varphi(\vx)\,\psi\bigr)(\vy_1...\vy_n) 
&= \sqrt{n+1}\int_{\RRR^3}d\vy\; \varphi(\vy-\vx) \, \psi\bigl(\vy_1...\vy_n,\vy\bigr)\label{aphidef1}\\
\bigl(a_\varphi^\dagger(\vx)\,\psi\bigr)(\vy_1...\vy_n) 
&= \frac{1}{\sqrt{n}} \sum_{k=1}^n \varphi(\vy_k-\vx)\,\psi\bigl(y\setminus \vy_k\bigr)\,.\label{aphidef2}
\end{align}
Also for the cut-off Hamiltonian $H^{\varphi}$, \eqref{phase} applies.

Alternatively, one can obtain a well-defined version of $H$ by means of an IBC, which is a condition on a wave function $\psi$ defined on a configuration space $\Q$ with boundaries that relates the values (or derivatives) of $\psi$ on the boundary of $\Q$ to the values of $\psi$ at suitable interior points of $\Q$. For particle creation, one takes $\Q$ as in \eqref{Qdef} and the IBC to relate boundary points of the $n$-particle sector to interior points in the $(n-1)$-particle sector, where the boundary configurations are those with two particles at the same location (``collision configurations''). We focus here on the spinless non-relativistic case based on the negative Laplacian operator as the free Hamiltonian of the bosons; for this case, IBCs were discussed in \cite{TT15a,TT15b,KS15,ibc2a,LS18,Lam18,co1} after previous work in \cite{Mosh51a,Mosh51b,Tho84,Yaf92, Tum04}. Bohmian trajectories associated with IBCs are defined in \cite{bohmibc}.
We find that \eqref{phase} applies again and more generally that, in many situations with IBCs, a generic choice of coefficients creates a time asymmetry. While it is true that for every such IBC, there is a ``time reversed IBC,'' the situation is perhaps not analogous to that of external fields because IBCs do not represent external fields that could be expected to transform in a non-trivial way. Be that as it may, IBCs lead to a novel type of time asymmetric Hamiltonians that seem worth studying, although we expect the true Hamiltonian of the universe to be time symmetric.

The remainder of this paper is organized as follows. 
In Section~\ref{sec:action} we present the main reasoning and results: that the action of time reversal on wave functions can be more than complex conjugation, that the action can be determined using Bohmian mechanics, that Hamiltonians with complex charges are time asymmetric if not all charges have equal or opposite phase, and that in the time asymmetric case, the probability current in the ground state can be nonzero. In Section~\ref{sec:emissiononly}, we ask whether there are IBCs that exclusively absorb particles, or ones that exclusively emit, and answer in the negative. In Section~\ref{sec:generic}, we show that general IBCs are generically time asymmetric. In Section~\ref{sec:gauge}, we show that changing all charges by $e^{i\theta}$ yields a physically equivalent Hamiltonian; up to physical equivalence, time reversal conjugates the wave function, and conjugates charges. In Section~\ref{sec:effective}, we characterize the behavior of complex charges by determining the effective potential with which they interact. In Section~\ref{sec:BC} we turn to ``ordinary'' boundary conditions (as opposed to IBCs), explain why such conditions, when local, are always time symmetric, and illustrate by means of an example how non-local boundary conditions can fail to be time symmetric. 
In Section~\ref{sec:conclusions}, we conclude.

\section{Action of $T$}
\label{sec:action}

In this section, we develop the basic questions, methods, and results step by step.

\subsection{Time Symmetry}

Although the expression \eqref{Hdef} for the Hamiltonian is UV divergent, let us ignore the divergence for a moment and work with \eqref{Hdef} pretending it was a self-adjoint operator.

We first claim that if all $g_j$ are real, then $H$ is time symmetric. In fact, we claim that in this case, the time reversal operator $T$ is nothing but complex conjugation, 
\be\label{Tdef1}
T(\psi)(\vy_1...\vy_n) = \psi^*(\vy_1...\vy_n)\,,
\ee
and one easily verifies (on the non-rigorous level) that
\be
TH = HT \,,
\ee
which then implies that
\be
T\Bigl( e^{-iHt} T(\psi) \Bigr) = e^{iHt} \psi \,,
\ee
expressing that $T(\psi)$ evolves backwards, proving time symmetry.

For $g=(g_1...g_N) \notin \RRR^N$, $H=H_g$ does not commute with conjugation. In fact,
\be\label{HgHg*}
(H_g\psi)^* = H_{g^*}(\psi^*)
\ee
with $g^*=(g_1^*...g_N^*)$, and
\be\label{expgg*}
\Bigl( e^{-iH_gt} (\psi^*) \Bigr)^* = e^{iH_{g^*}t} \psi \,,
\ee
so $H_{g^*}$ takes the place of $H_g$, which is a different operator. This suggests that $H_g$ is time asymmetric; however, this does not necessarily follow, for the following reason. We might re-define $T$ by
\be\label{Tdef2}
T(\psi)(\vy_1...\vy_n) =  e^{-i2\theta n} \psi^*(\vy_1...\vy_n)
\ee
for a real constant $\theta$. This means that in addition to complex conjugation, we multiply by a function on configuration space $\Q=\cup_n \Q_n$ that is constant on each sector but depends on the particle number $n$. Now, if 
\be\label{gdef}
g_j = e^{i\theta} \tilde g_j \text{ with } \tilde g_j \in\RRR
\ee
(so that any $g_i,g_j$ have either equal or opposite phases depending on whether $\tilde g_i, \tilde g_j$ have equal or opposite signs),
then
\be\label{HgHg}
TH_g = H_g T\,,
\ee
as
\begin{align}
(TH_g\psi)(y) 
 &= e^{-i2\theta n} H_\free \psi^*(y) + e^{-i\theta(2n+1)} 
     \sqrt{n+1} \sum_{j=1}^N \tilde g_j 
     \, \psi^*(y,\vx_j) \nonumber\\
 & \quad + \: e^{-i\theta(2n-1)} \frac{1}{\sqrt{n}} \sum_{j=1}^N
   \sum_{k=1}^n \tilde g_j \, \delta^3(\vy_k-\vx_j) 
   \, \psi^*(y\setminus \vy_k) \label{calc1}\\
 & = (H_gT\psi)(y) \,.
\end{align}
Now \eqref{HgHg} implies that 
\be\label{TUT}
T e^{-iH_gt} T\psi = e^{iH_gt}\psi\,,
\ee
so $H_g$ is time symmetric if $T$ given by \eqref{Tdef2} represents time reversal. Note that $T$ is anti-unitary (as it is the composition of the anti-unitary conjugation and the unitary multiplication operator $e^{-i2\theta n}$), and that 
\be\label{T2T}
T^2=T\,.
\ee

So the question arises, how do we know whether \eqref{Tdef1} or \eqref{Tdef2} or something else is the correct action of time reversal for $H_g$? 

We will answer this question using Bohmian mechanics \cite{DT} as follows. Bohmian mechanics provides, for a given $\psi$ and $H$, a stochastic process $(Q_t)_{t\in\RRR}$ in configuration space with the Markov property and the property that for every $t\in\RRR$, $Q_t$ has distribution $|\psi_t|^2$. In the Bohmian theory, one can ask whether the process for $\psi^*$, or that for $T\psi$, is the time reverse of that for $\psi$. If that is so, then that justifies the statement that $\psi\to \psi^*$, or $\psi\to T\psi$, represents time reversal. For example, one finds that this is so for $\psi^*$ if the Hamiltonian is of the form $-\Delta + V$ on $\RRR^n$. In order to define the Bohmian process for our QFT, we first need a well-defined Hamiltonian, for which we consider two options: UV cut-off and IBC.

\subsection{Well-Defined Hamiltonian}

We have already defined the Hamiltonian $H^\varphi_g$ with cut-off function $\varphi$, which we assume to be real-valued. We remark that for $H_g^\varphi$, \eqref{HgHg*} is still correct (which yields time symmetry for real charges), and so is the commutation relation \eqref{HgHg} with $T$ given by \eqref{Tdef2} and $g$ by \eqref{gdef}, as follows from a calculation analogous to \eqref{calc1}.

The IBC technique rigorously defines a Hamiltonian $H^{IBC}$ without the need for a UV cut-off. The Hamiltonian is defined \cite{ibc2a} on a domain of wave functions in $\Hilbert$ satisfying the following IBC: For  $y=(\vy_1...\vy_n)\in(\RRR^3\setminus\{\vx_1...\vx_N\})^n$, $k\leq n$, and $j\in\{1...N\}$,
\be\label{IBC1}
\lim_{\vy_k\to \vx_j} \,|\vy_k-\vx_j| \,\psi(y)
= -\tfrac{mg_j^*}{2\pi\sqrt{n}}\, \psi(y\setminus \vy_k)\,.
\ee
The corresponding Hamiltonian is 
\begin{align}
(H^{IBC}_g\psi)(y) 
&=-\tfrac{1}{2m} \sum_{k=1}^n \nabla^2_{\vy_k}\psi
\nonumber\\
& + \frac{\sqrt{n+1}}{4\pi} \sum_{j=1}^N g_j \int\limits_{\SSS^2} \!\! d^2\vomega \, \lim_{r\searrow 0} \partial_r \Bigl[ r \psi\bigl(y,\vx_j+r\vomega \bigr) \Bigr]\nonumber\\
& +\: \frac{1}{\sqrt{n}} \sum_{j=1}^N g_j^*\sum_{k=1}^n  \delta^3(\vy_k-\vx_j)\,\psi\bigl(y\setminus \vy_k\bigr) \,.\label{Hdef1}
\end{align}
The self-adjointness of this Hamiltonian is proved in \cite{ibc2a} (see \cite{LS18,Lam18,Sch18} for further developments of similar proofs). Let us check \eqref{HgHg*}: If $\psi$ satisfies the IBC \eqref{IBC1} with given $g=(g_1...g_N)$, then $\psi^*$ satisfies it with $g^*$ instead of $g$. In fact (as becomes clear from \cite{ibc2a}), if $\psi$ lies in the domain of $H^{IBC}_g$, then $\psi^*$ lies in that of $H^{IBC}_{g^*}$, and, as visible from the explicit form \eqref{Hdef1} of $H^{IBC}_g$, \eqref{HgHg*} holds. Likewise for \eqref{HgHg}: If $\psi$ satisfies the IBC \eqref{IBC1}, $g$ satisfies \eqref{gdef}, and $T$ is given by \eqref{Tdef2}, then $T\psi$ satisfies the IBC \eqref{IBC1} as well (because $T\psi(y) = e^{-i2\theta n} \psi^*(y)$ while $g_j^* T\psi(y\setminus \vy_k) =  e^{-i\theta} \tilde g_j e^{-i2\theta(n-1)}  \psi^*(y\setminus \vy_k)$, so $e^{-i2\theta n}$ times the conjugate of \eqref{IBC1} yields \eqref{IBC1} for $T\psi$). Then, both $H^{IBC}_gT$ and $TH_g^{IBC}$ yield the right-hand side of \eqref{Hdef1} with $\psi$ replaced by $\psi^*$ and each term multiplied by the same phase factor as the corresponding term in \eqref{calc1}, thus proving \eqref{HgHg} for \eqref{Hdef1}.

\subsection{Bohmian Trajectories}

Now we can introduce Bohmian trajectories for $H_g^\varphi$ \cite{DGTZ04,DGTZ05b} and $H_g^{IBC}$ \cite{bohmibc}. Except for particle creation and annihilation, the actual configuration $Q_t\in \Q$ moves according to Bohm's equation of motion,
\be\label{Bohm}
\frac{dQ_t}{dt}= v^{\psi_t}(Q_t) = \tfrac{1}{m}\, \Im \frac{\nabla\psi_t}{\psi_t}(Q_t)
\ee
with $\psi_t=e^{-iHt}\psi$. Particle creation and annihilation corresponds to a jump of $Q_t$ to the next higher or lower sector. 

In the version of the theory with UV cut-off $\varphi$, all of these jumps occur stochastically. During any time interval of length $dt$, a particle gets created in the volume $d^3\vy'$ around $\vY' \in \RRR^3$, i.e., the configuration $Q_t=Y=(\vY_1...\vY_n)$ jumps to $Y'=(\vY_1...\vY_k,\vY', \vY_{k+1}...\vY_n)$, with probability
\be\label{jumprate1}
\sigma(Y\to Y',t) \,d^3\vy'\, dt = 
\tfrac{2}{\sqrt{n+1}} \frac{\max\bigl\{0, \Im \bigl[\psi_t^*(Y') \sum_{j=1}^N g_j^* \varphi(\vY'-\vx_j) \, \psi_t(Y) \bigr] \bigr\}}{|\psi_t(Y)|^2} d^3\vy' \, dt \,.
\ee
(That is, this is the conditional probability, given $Q_t$.)
Likewise, during any time interval of length $dt$, particle $k$ gets annihilated, i.e., the configuration $Q_t=Y=(\vY_1...\vY_n)$ jumps to $Y'=Y\setminus \vY_k$, with probability
\be\label{jumprate2}
\sigma(Y\to Y',t) \, dt = 
\tfrac{2}{\sqrt{n}} \frac{\max\bigl\{0, \Im \bigl[\psi_t^*(Y') \sum_{j=1}^N  g_j \varphi(\vY_k-\vx_j) \, \psi_t(Y) \bigr] \bigr\}}{|\psi_t(Y)|^2} dt \,.
\ee

In the version of the theory with IBC, as soon as one of the $y$-particles reaches one of the $x$-particles, it gets annihilated. Conversely, during any time interval of length $dt$, a particle gets emitted at $\vx_j$ in the direction $\vomega$, i.e., $Q_t=Y$ jumps to $Y'=(\vY_1...\vY_k,\vx_j+0\vomega, \vY_{k+1}...\vY_n)$ (see \cite{bohmibc} for more detail), with probability
\be\label{jumprate3}
\sigma(Y\to Y',t) \,d^2\vomega\, dt = 
\tfrac{1}{m} \lim_{r\to 0} \frac{\max\bigl\{0, \Im \bigl[r^2 \, \psi_t^*(Y'_r) \, \partial_r \psi_t(Y'_r) \bigr] \bigr\}}{|\psi_t(Y)|^2} d^2\vomega \, dt \,,
\ee
where $Y'_r := (\vY_1...\vY_k,\vx_j+r\vomega, \vY_{k+1}...\vY_n)$. From the point $Y'$ on the boundary of the $(n+1)$-particle sector of $\Q$, $Q_t$ moves into the interior according to \eqref{Bohm} \cite{TT15b,bohmibc}. 

In both versions of the theory, with cut-off or with IBC, it follows from the defining laws that if $Q_t$ is $|\psi(t)|^2$ distributed for $t=0$, then $Q_t$ is $|\psi(t)|^2$ distributed also for $t>0$ \cite{DGTZ05b,bohmibc}. One thus obtains a stochastic process $(Q_t^\psi)_{t\geq 0}$ associated with every (normalized) initial wave function $\psi$. This process can naturally be defined also for negative times $t$ as follows. For any $\tau\in\RRR$, a process $(Q_t^{\psi})_{t\geq \tau}$ starting at time $\tau$ with a $|\psi_\tau|^2$-distributed configuration can be defined correspondingly and has, as its marginal for $t\geq \tau'>\tau$, a process that is equal in distribution to the process $(Q_t^\psi)_{t\geq \tau'}$ associated with $\tau'$. Thus, all of these processes for any $\tau\in\RRR$ fit together to form a single process $(Q_t^\psi)_{t\in\RRR}$.

\subsection{Proof of Time Symmetry}

Let us apply the Bohmian criterion to the question of how to represent time reversal. Suppose that all $g_j$ have equal or opposite phases, $g_i^*g_j\in\RRR$. It turns out that in both versions, the Bohmian theory is actually time symmetric if $T$ as in \eqref{Tdef2} represents time reversal, but not for mere conjugation. We conclude that \eqref{Tdef2} is the correct representation of time reversal for $g$ as in \eqref{gdef}, and $\psi \to \psi^*$ is not.

\bigskip

Indeed, in order to verify the symmetry of the Bohmian process, we begin by noting that the reverse $\hat Q_t=Q_{-t}$ of a Markovian jump process is again a Markovian jump process. Its velocities are reversed, $\hat v(q,t) = -v(q,-t)$; its distribution at time $t$ equals that of $Q$ at time $-t$, $\hat\rho(dq,t)=\rho(dq,t)$; in our case, $\rho(dq,t) = |\psi_t(q)|^2 dq$. And its jump rates can be computed as follows: the amount of probability moved from the infinitesimal volume $dq$ to $dq'$ during $[t,t+dt]$ in the process $Q$ is $\sigma(q\to dq',t) \, \rho(dq,t) \, dt$. The same amount is moved in the reversed process $\hat Q$ from $dq'$ to $dq$ during $[-t-dt,-t]$, so its jump rate $\hat\sigma$ obeys
\be\label{amount}
\hat\sigma(q\to dq',t) \, \hat\rho(dq,t)\, dt
= \sigma(q'\to dq,-t) \, \rho(dq',-t) \, dt\,.
\ee
Hence, since equivariance takes care of the right $\hat\rho$, the condition for the time symmetry of a Bohmian theory with jumps, given a proposed action $T$ of time reversal, is that firstly \eqref{HgHg} holds, secondly
\be\label{reversev}
v^{T\psi}(q) = -v^{\psi}(q)\,,
\ee
and thirdly,
\be\label{reversesigma}
\sigma^{T\psi}(q\to dq') 
= \frac{\sigma^{\psi}(q'\to dq) \, |\psi(q')|^2 \, dq'}
{|T\psi(q)|^2 \,dq} \,.
\ee

Concerning \eqref{HgHg}, we have already seen that it holds for $T$ as in \eqref{Tdef2} but not in general for conjugation.

Concerning \eqref{reversev}, both conjugation and $T$ as in \eqref{Tdef2} will reverse velocities according to Bohm's equation of motion \eqref{Bohm} because the the imaginary part changes sign under conjugation, while the factor $e^{-i2\theta n}$ is locally constant and thus does not affect the gradient of $\psi$ except for a phase factor that cancels out of \eqref{Bohm}. 

Concerning \eqref{reversesigma} in the case with cut-off $\varphi$, note first that for particle creation, $dY'= dY \, d\vy'$; note also that for any $q,q'$, only one of the jumps $q\to q'$ and $q'\to q$ is allowed at any time, that is, the other has rate 0. Thus, to obtain \eqref{reversesigma}, it suffices that under time reversal and exchange $q \leftrightarrow q'$, $\Im[...]$ in \eqref{jumprate1} and \eqref{jumprate2} changes sign and hence that the square bracket gets conjugated. Mere conjugation of $\psi$ does not achieve that because it does not affect $g_j$, whereas conjugation of the square bracket in \eqref{jumprate1} or \eqref{jumprate2} also conjugates $g$. But $T$ as in \eqref{Tdef2} does achieve that because the additional phases introduced by $T$ into $\psi(Y)$ and $\psi(Y')$ differ by $e^{\pm i2\theta}$ which, assuming $g$ as in \eqref{gdef}, will interchange $g_j$ and $g_j^*$. Thus, \eqref{reversesigma} is satisfied.

Concerning \eqref{reversesigma} in the case with IBC, it is easiest to go back to \eqref{amount} and compare the amounts of probability transported from $q$ to $q'$. The amount of probability flowing per time into the surface element $d^2\vomega \, dY$ with the $(k+1)$-st $y$-particle hitting $\vx_j$ is the flux of the probability current into that surface element, which is
\be
\lim_{r\to 0} \max \Bigl\{ 0, -\tfrac{1}{m} \Im \bigl[ r^2 \, \psi^*(Y'_r) \, \partial_r \psi(Y'_r) \bigr] \Bigr\} \,.
\ee
So, we need that this quantity coincides with $\hat\rho(Y) \, \hat\sigma(Y \to Y')$, which it does for both conjugation and $T$ as in \eqref{Tdef2} because both will conjugate the square bracket as the factor of $e^{-i2\theta (n+1)}$ cancels out.\hfill$\square$

\subsection{Time Asymmetry}

The Bohmian criterion also allows us to prove that whenever the charges do not have equal or opposite phases, the theory violates time reversal symmetry. Without the Bohmian criterion, it would remain unclear how to prove such a violation because it would be unclear which anti-unitary operator $T$ is the correct representation of time reversal. In the Bohmian framework, we can simply prove that there is no operator for the role of $T$ that would make the theory time symmetric.

\bigskip

Indeed, assuming that none of the $g_j$ vanishes and that they do not have equal or opposite phases, we will identify restrictions on which kind of operator $\widetilde T$ might represent time reversal, and then to show that $H_g^\varphi$ and $H_g^{IBC}$ commute with none of these operators. We take for granted that $\widetilde T$ is real-linear and norm-preserving. So suppose that for every (normalized) wave function $\psi$ there is another one, $\widetilde T \psi$, such that $Q^{\widetilde T\psi}_t=Q^\psi_{-t}$ in distribution. Since the Bohmian velocity field $v^\psi$ is the gradient of the phase of $\psi$, $\widetilde T \psi$ must have the opposite phase of $\psi$ up to addition of a function $\theta(q)$ on $\Q$ that has gradient 0 and thus is locally constant (i.e., constant on every sector). Since the probability distribution of $Q_0$ is simultaneously given by $|\psi|^2$ and by $|\widetilde T\psi|^2$, $\psi$ and $\widetilde T \psi$ must have equal modulus. So,
\be\label{asym}
(\widetilde T \psi)(\vy_1...\vy_n) = e^{i\theta(n)}\psi^*(\vy_1...\vy_n)\,.
\ee

We now show that if $H_g^\varphi$ commutes with one such $\widetilde T$, then the $g_j$ have equal or opposite phases. We find that
\begin{align}
 (H_g^\varphi \widetilde T \psi)(y) 
 &~=~ e^{i\theta(n)} H_\free \psi^*(y) 
  + \sqrt{n+1} \sum_{j=1}^N  g_j \int d\vy\, 
  \varphi(\vy-\vx_j)\,e^{i\theta(n+1)}\,\psi^*(y,\vy) \nonumber\\
 &\quad + \frac{1}{\sqrt{n}} \sum_{j=1}^N \sum_{k=1}^n  g^*_j \,
 \varphi(\vy_k-\vx_j) e^{i\theta(n-1)} \,\psi^*(y\setminus \vy_k) \\
 (\widetilde T H_g^\varphi \psi)(y) 
 &~=~ e^{i\theta(n)} H_\free \psi^*(y) 
  + e^{i\theta (n)} \sqrt{n+1} \sum_{j=1}^N   g^*_j \int d\vy\, 
  \varphi(\vy-\vx_j) \, \psi^*(y,\vy) \nonumber\\
 &\quad + e^{i\theta (n) }\, \frac{1}{\sqrt{n}} \sum_{j=1}^N 
  \sum_{k=1}^n  g_j \, \varphi(\vy_k-\vx_j) \, 
  \psi^*(y\setminus \vy_k) 
\end{align}
Let us compare the second term in each equation, as well as the third. If these are to be equal for all $\psi$, we need that 
\begin{align}
g_j e^{i\theta(n+1)} &= e^{i\theta(n)} g^*_j \quad \text{and}\\
g^*_j e^{i\theta(n-1)} &= e^{i\theta(n)} g_j \label{n-1n}
\end{align}
for all $j$ and all $n$, which is equivalent to
\be
\frac{g_j}{g^*_j} = e^{i\theta(n)-\theta(n+1)}
\ee
for all $j$ and all $n$, so $g_i/g^*_i = g_j/g^*_j$ or $g_i g^*_j = g^*_i g_j$ or $g^*_i g_j \in \RRR$. This completes the proof.\hfill$\square$

\bigskip

We now show that if $H_g^{IBC}$ commutes with one such $\widetilde T$, then the $g_j$ have equal or opposite phases. In order to obtain \eqref{TUT}, the commutation relation $TH_g=H_gT$ must be understood as including that for $\psi$ in the domain of $H_g$, also $T\psi$ lies in the domain. We show that if the IBC \eqref{IBC1} for $\psi$ implies that for $\widetilde T\psi$, then the $g_j$ have equal or opposite phases. Comparing the conjugate of \eqref{IBC1},
\be\label{IBC1*}
\lim_{\vy_k\to \vx_j} \,|\vy_k-\vx_j| \,\psi^*(y)
= -\tfrac{mg_j}{2\pi\sqrt{n}}\, \psi^*(y\setminus \vy_k)\,,
\ee
to the IBC for $\widetilde T\psi$,
\be\label{IBC1T}
\lim_{\vy_k\to \vx_j} \,|\vy_k-\vx_j| \,e^{i\theta(n)}\psi^*(y)
= -\tfrac{mg^*_j}{2\pi\sqrt{n}}\, e^{i\theta(n-1)}\psi^*(y\setminus \vy_k)\,,
\ee
shows that, for all $j$ and all $n$, 
\be
g_j\, e^{i\theta(n)} = g_j^* \, e^{i\theta(n-1)}\,,
\ee
which is the same as \eqref{n-1n} and thus implies $g_i^* g_j \in \RRR$. This completes the proof.\hfill$\square$

\subsection{Nonzero Current in the Ground State}

It is a situation familiar from many examples that the ground state of a quantum system has vanishing current (so that the Bohmian particles do not move). This situation is, in fact, related to time symmetry: If a Hamiltonian is time symmetric then any non-degenerate eigenstate must be invariant under time reversal. In particular, if time reversal is given by complex conjugation, then any non-degenerate eigenstate must be real up to a global phase factor. Since currents change sign under time reversal, but are the same because the time-reversed state is the same up to a global phase, the currents must vanish. Correspondingly, we should expect this property to fail for time \emph{asymmetric} Hamiltonians. In this section, we confirm for an example of a time asymmetric Hamiltonian that the ground state is non-real and exhibits non-zero current.

\begin{figure}[h]
\begin{center}
\includegraphics[width=0.7\textwidth]{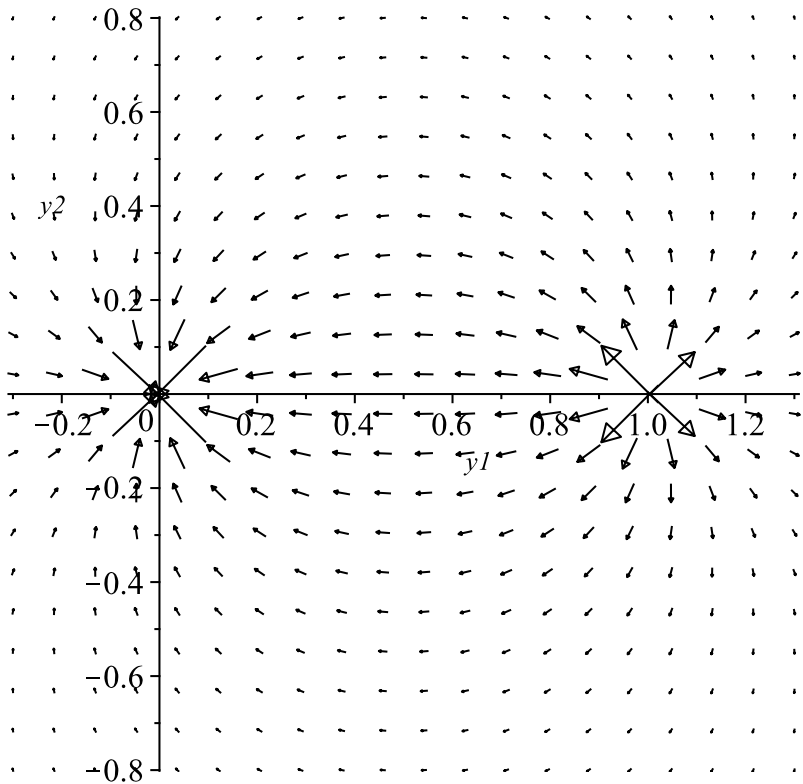}
\end{center}
\caption{Plot of two components of the current vector field $\vj(\vy)$ in the 1-particle sector of configuration space for the ground state \eqref{groundstate} of $H_g^{IBC}$ with $N=2$ charges located at the origin and $(1,0,0)$; the charges have phases that are neither equal nor opposite, and $\sqrt{2mE_0}/\hbar$ is taken to be $1/10|\vx_1-\vx_2|$.}
\label{fig:current}
\end{figure}

For $H_g^{IBC}$ with $E_0>0$, the ground state is (with factors of $\hbar$ made explicit)
\be\label{groundstate}
\psi_{\min}(\vy_1,\ldots,\vy_n)= \mathcal{N}\frac{(-m)^n}{(2\pi\hbar^2)^n\sqrt{n!}}
\prod_{k=1}^n \psi_1(\vy_k)\,,
\ee 
where $\mathcal{N}$ is a normalizing constant and $\psi_1$ an abbreviation for 
\be
\psi_1(\vy)=\sum_{j=1}^N g_j^* \frac{e^{-\sqrt{2mE_0}|\vy-\vx_j|/\hbar}}{|\vy-\vx_j|}\,.
\ee
This formula agrees with what we found in \cite{TT15a,TT15b} for the case $g_i=g_j\in\RRR$.

In $\psi_{\min}$, each sector is a tensor power of $\psi_1$, so each boson has the same wave function $\psi_1$, and if (and only if) the charges $g_j$ do not have equal or opposite phases, then $\psi_1$ does not have constant phase (since the phase of $\psi_1(\vy)$ is close to that of $g_j^*$ when $\vy$ is close to $\vx_j$). As a consequence, $\psi_{\min}$ does not have constant phase, and since the current is $m^{-1}|\psi|^2$ times the gradient of the phase, the current is nonzero. 

\begin{figure}[h]
\begin{center}
\includegraphics[width=0.7\textwidth]{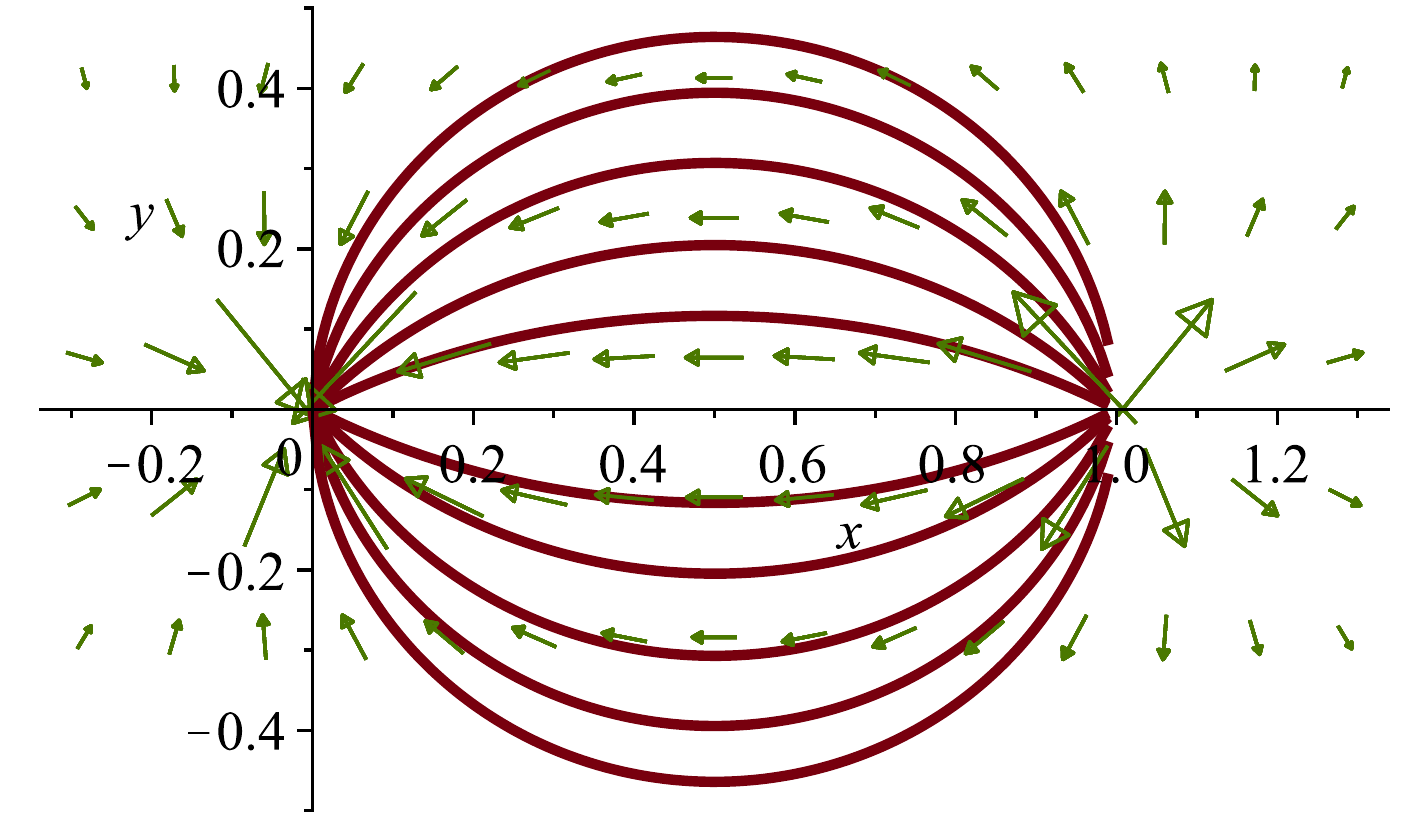}
\end{center}
\caption{A selection of 10 integral curves (Bohmian trajectories) of the vector field shown in Figure~\ref{fig:current}.} 
\label{fig:current2}
\end{figure}

Here is the explicit expression for the current. In the $n$-sector, the current has $3n$ components, of which those for particle $k$ are just
\be
\vj^{\psi_{\min}}_k(\vy_1...\vy_n) \propto \vj^{\psi_1}(\vy_k) \prod_{\ell\neq k} |\psi_1(\vy_\ell)|^2\,,
\ee
so each $y$-particle moves along an integral curve of $\vj^{\psi_1}$. The vector field $\vj^{\psi_1}$ for $N=2$ charges is depicted in Figure~\ref{fig:current}, and some of its integral curves in Figure~\ref{fig:current2}. Its explicit form is
\be
\vj^{\psi_1}(\vy) = \tfrac{\hbar}{m} \sum_{\substack{i,j=1\\i\neq j}}^N \Im[g_i^*g_j] \frac{e^{-\alpha r_i} e^{-\alpha r_j}}{r_i r_j}\Bigl(\alpha+\frac{1}{r_j} \Bigr)\ve_j
\ee
with the abbreviations $\alpha=\sqrt{2mE_0}/\hbar$, $r_i=|\vy-\vx_i|$, and $\ve_j= (\vy-\vx_i)/r_i$ the unit vector pointing from $\vx_i$ towards $\vy$.

In the case of two charges, if $\Im[g_1^*g_2]>0$, then $\vj^{\psi_1}$ is pointing away from charge 2 and pointing towards charge 1, and the Bohmian process looks as follows: At random times (in fact, at a constant rate), bosons get created at charge 2 and emitted in a random direction, move along a flow line of $\vj^{\psi_1}$, and finally hit charge 1, where they get annihilated. Each $y$-particle follows the Bohmian 1-particle velocity field $\vj^{\psi_1}/|\psi_1|^2$, independently of the other $y$-particles; likewise, the initial direction of any one $y$-particle is independent of the others. If the last $y$-particle gets annihilated, then the process will wait in the vacuum configuration until the next $y$-particle gets created.

This concludes the presentation of our basic results. The next three sections provide a deeper investigation of some aspects of the topic, the most remarkable of which is perhaps the observation in Section~\ref{sec:equiv} that the Hamiltonian $H_{e^{i\theta}g}$ is physically equivalent to $H_g$.

\section{Every IBC Involves Both, Emission and Absorption}
\label{sec:emissiononly}

The last example inspires us to ask: Could there be IBCs that enable exclusively emission (i.e., jumps from the $n$-sector to the $n+1$-sector), or exclusively absorption (i.e., jumps from the $n+1$-sector to the $n$-sector)? After all, in the last example, only emissions occured at $\vx_2$, and only absorptions at $\vx_1$. Of course, this was for a special wave function, the ground state, and our question in this section is whether an IBC could be set up in such a way that this happens for \emph{every} wave function. For time symmetric $H$, of course, this cannot happen because the time reverse of an emission process is an absorption, and that would occur with the same $H$ and a different wave function $T\psi$.

For a time asymmetric $H$, however, one could perhaps imagine that a purely emitting boundary is the time reverse of a purely absorbing boundary and corresponds to a different Hamiltonian. Yet, we now show that the answer to our question is negative.

For simplicity, we focus on boundaries of codimension 1.\footnote{The boundary at which one of the $y$-particles meets $\vx_j$ is in a sense not so different because in spherical coordinates centered at $\vx_j$, the boundary corresponds to $r=0$, which looks like a codimension-1 surface in these coordinates.} A general IBC will be of the form 
\be\label{emissionIBC}
(\alpha+\beta \partial_n)\psi(q') = \psi(q)\,,
\ee
where $\alpha$ and $\beta$ are complex constants, $q'$ is a point on the boundary (in, say, the $n+1$-sector), $q$ is the associated interior point (in the $n$-sector), and $\partial_n$ is the derivative in the direction normal to the boundary. The current into the boundary is
\be
j_n(q') = \tfrac{\hbar}{m} \, \Im \bigl[ \psi^*(q') \partial_n \psi(q') \bigr] \,.
\ee
The question is whether the IBC \eqref{emissionIBC} could be designed in such a way that $j_n$ is always $\neq 0$ (or perhaps always $\leq 0$). We will show that the answer is ``no''; more precisely, that for given $\alpha,\beta$, and $\psi(q)$ with $(\alpha,\beta)\neq (0,0)$ and $\psi(q)\neq 0$, there exist values $u$ and $v$ for $\psi(q')$ and $\partial_n \psi(q')$ that make $j_n(q')$ positive and others $\tilde u$ and $\tilde v$ that make $j_n(q')$ negative.

\bigskip

Indeed, if $\beta=0$, then the IBC \eqref{emissionIBC} is equivalent to $u=\psi(q)/\alpha$, so
\be
j_n(q') = \tfrac{\hbar}{m} \, \Im [\tfrac{\psi^*(q)}{\alpha^*}v]\,,
\ee
which can be made positive or negative by varying $v$. If, however, $\beta \neq 0$, then the IBC \eqref{emissionIBC} is equivalent to $v= (\psi(q)-\alpha u)/\beta$, so, writing $u=re^{i\varphi}$ and $\psi(q)/\beta=s e^{i\chi}$ in polar coordinates (with $r,s\geq 0$ and $\varphi,\chi\in\RRR$), 
\begin{align}
\tfrac{m}{\hbar}\, j_n(q')
&= \Im[u^* \tfrac{\psi(q)}{\beta}] - |u|^2 \, \Im [\tfrac{\alpha}{\beta}] \\
&= rs \sin(\chi-\varphi) - r^2 \Im [\tfrac{\alpha}{\beta}] =: f(r,\varphi)\,.
\end{align}
Since $f(0,\varphi)=0$ and $\frac{\partial f}{\partial r} (0,\varphi)= s \sin(\chi-\varphi)$, it is clear that $\frac{\partial f}{\partial r} (0,\varphi)$ (and thus, for small values of $r$, $f(r,\varphi)$) can be made positive (negative) by choosing $\varphi$ so that $\sin(\chi-\varphi)$ is positive (negative).\hfill$\square$

\bigskip

Readers may wonder how this result can be compatible with the existence of absorbing boundary conditions such as $\partial_n \psi = i\kappa \psi$ with $\kappa>0$ \cite{detect-rule}. The answer is that this absorbing boundary condition is included in \eqref{emissionIBC} for $\psi(q)=0$, which leads to $\tfrac{\partial f}{\partial r}=0$, whereas our argument assumed that $\psi(q)\neq 0$, which of course will happen for some wave functions.

\section{General IBCs Are Generically Time-Asymmetric}
\label{sec:generic}

We have already seen that a generic choice of $g\in \CCC^N$ makes the IBC \eqref{IBC1} time asymmetric. However, the IBC \eqref{IBC1} is not the most general one possible for particle creation, as discussed in \cite{TT15a,TT15b,ibc2a} (see also \cite{co1}).  In this section, we consider general IBCs and show that they, too, are generically time asymmetric.

With the abbreviations  \cite{ibc2a}
\begin{align}
(B_{j}\psi)(y)&=-\tfrac{\sqrt{n+1}}{2m}\int\limits_{\SSS^2} \!\! d^2\vomega \,\lim_{r\searrow 0} \,r \,\psi(y,\vx_j + r\vomega)\\
(A_{j}\psi)(y)&= \tfrac{\sqrt{n+1}}{4\pi} \int\limits_{\SSS^2} \!\! d^2\vomega \, \lim_{r\searrow 0} \partial_r \Bigl[ r \psi\bigl(y,\vx_j+r\vomega \bigr) \Bigr]\,,
\end{align}
the IBC \eqref{IBC1} reads
\be
\tfrac{1}{g_j^*} \, B_j\psi(y) = \psi(y)\,,
\ee
and the Hamiltonian \eqref{Hdef1}, when acting on the set of configurations $y$ with $\vy_k\neq \vx_j$ for all $j,k$, has the form
\be
H\psi(y) = H_\free\psi(y) + \sum_{j=1}^N g_j \, A_{j}\psi(y) \,.
\ee

The most general IBC is obtained by replacing
\begin{align}
\frac{1}{g_j^*}B_{j} &\to \hat B_{j}:= e^{i\theta_j} (\alpha_j B_{j} + \beta_j A_{j})\\
g_j \, A_{j} &\to \hat A_{j}:= e^{i\theta_j} (\gamma_j B_{j} + \delta_j A_{j})
\end{align}
with real $\alpha_j, \beta_j, \gamma_j, \delta_j$ obeying
\be
\alpha_j \delta_j - \gamma_j \beta_j =1\,,
\ee
which yields
\be\label{IBC2}
\hat B_{j}\psi(y) = \psi(y)
\ee
as the IBC and
\be
H\psi(y) = H_\free\psi(y) + \sum_{j=1}^N \hat A_{j}\psi
\ee
as the Hamiltonian on the set of configurations $y$ with $\vy_k\neq \vx_j$ for all $j,k$.

For $\widetilde T$ as in \eqref{asym}, one finds that each of the relations $\widetilde T \hat B_j \widetilde T= \hat B_j$ and $\widetilde T  \hat A_j \widetilde T= \hat A_j$ holds iff $e^{i\theta(n+1)-i\theta(n)}=e^{-i2\theta_j}$ or, equivalently, 
\be\label{condition}
\theta(n+1)-\theta(n)=-2\theta_j\text{ mod }2\pi\,.
\ee
If the phases $e^{i\theta_j}$ are mutually either equal or opposite (i.e., $\theta_i=\theta_j$ mod $\pi$), then this condition can be satisfied by choosing $\theta(n) = -2n\theta_j$.
Conversely, the condition \eqref{condition} cannot be satisfied for all $j$ simultaneously if the $\theta_j$ are different mod $\pi$, i.e., if the phases $e^{i\theta_j}$ are neither equal nor opposite. 

To sum up, {\it the general IBC \eqref{IBC2} is time symmetric iff all phases $e^{i\theta_j}$ are equal or opposite; thus, for generic $\theta_j$ they are time asymmetric.}

\section{Time-Reversed Hamiltonian}
\label{sec:gauge}

For any given Hamiltonian $H$ and any given action $T$ of time reversal, one can define the \emph{time-reversed Hamiltonian} by
\be
H^\rev = THT\,.
\ee
Then $H$ is time symmetric if and only if $H=H^\rev$. In our case, the situation is a bit more complicated for two reasons, first because of the several possibilities for what $T$ could be and second because some Hamiltonians are physically equivalent to others, as we explain now.

\subsection{Physically Equivalent Hamiltonians}
\label{sec:equiv}

It is widely accepted that adding a constant (i.e., a multiple of the identity) to the Hamiltonian, $H+E$, does not correspond to a physical change. We will now argue that, for $g\in \CCC^N$ and $\theta\in\RRR$, replacing $H_g$ by $H_{e^{i\theta}g}$ is no physical change either. In orthodox quantum mechanics, one could argue that there is no physical difference between two situations if the distributions of outcomes for measurements of arbitrary observables are the same (although such an argument does not seem fully convincing as there might be limitations to knowledge \cite{CT16}). In Bohmian mechanics, one can argue that there is no physical difference between two situations if the possible trajectories and their probabilities are the same. We will verify this Bohmian criterion. (It then follows that also the distributions of outcomes of arbitrary experiments are the same.)

We will proceed as follows, carrying out the reasoning for both $H_g=H_g^\varphi$ and $H_g=H_g^{IBC}$. We define the unitary operator $U$ by
\be\label{Udef}
(U\psi)(\vy_1...\vy_n)= e^{-i\theta n}\psi(\vy_1...\vy_n)\,,
\ee
then we show that if $\psi$ evolves with $H_g$, $\psi_t = e^{-iH_gt} \psi$, then $U\psi$ evolves with $H_{e^{i\theta}g}$, $U\psi_t=e^{-iH_{e^{i\theta}g}t}U\psi$. Finally, we show that the Bohmian process $(Q_t)_{t\in\RRR}$ is the same for $\psi$ and $U\psi$.

Indeed, the statement about the time evolution follows from 
\be\label{HUHU}
H_g = U^{-1} H_{e^{i\theta}g} U \,.
\ee
For $H_g^\varphi$, this easily follows from the facts that $U$ commutes with $H_\free$, that $a_\varphi(\vx)U=e^{-i\theta}Ua_\varphi(\vx)$ (as one sees easily from the definition \eqref{aphidef1} of $a_\varphi$), and that $a_\varphi^\dagger(\vx)U = e^{i\theta}Ua_\varphi^\dagger(\vx)$.

For $H_g^{IBC}$, one reads off of the IBC \eqref{IBC1} that if $\psi$ satisfies the IBC with $g$ then $U\psi$ satisfies it with $e^{i\theta}g$. Likewise, from the definition \eqref{Hdef1} of $H_g^{IBC}$ one reads off that \eqref{HUHU} holds.

Concerning the Bohmian process, the factor $e^{-i\theta n}$ cancels out of Bohm's equation of motion; in the $\Im[...]$ expression in the jump rate formula \eqref{jumprate1} with $\psi$ replaced by $U\psi$ and $g$ replaced by $e^{i\theta}$, the $\psi_t^*(Y')$ contributes a factor $e^{i\theta (n+1)}$, the $g_j^*$ a factor $e^{-i\theta}$, and the $\psi_t(Y)$ a factor $e^{-i\theta n}$, so that these phase factors cancel. The same happens in the other jump rate formulas \eqref{jumprate2} and \eqref{jumprate3}.\hfill$\square$

\bigskip

To put things differently, we define that pairs $(g,\psi)$ and $(g',\psi')$ are \emph{equivalent} iff there is $\theta\in\RRR$ such that $g'=e^{i\theta}g$ and $\psi'=U\psi$. Then equivalent pairs have the same Bohmian process. We therefore regard them as representing the same physical reality. This equivalence relation is similar to a change of gauge, where one considers a pair $(A_\mu,\psi)$ of a gauge connection and a wave functions gauge equivalent to the pair $(A_\mu + \partial_\mu f(\vx,t), e^{if}\psi)$.

\subsection{Time Reversal as Conjugating the Charge}

If we tacitly understand that the appropriate change $U$ has been applied to $\psi$ when changing $g$, we can also talk about equivalence between Hamiltonians, which would appropriately be called \emph{physical equivalence}.
Thus, we can say that $H_g$ is time symmetric iff it is physically equivalent to an $H_{\tilde g}$ with real $\tilde g_j$. 

More generally, if we consider Hamiltonians only up to physical equivalence, then time reversal for $g$ with equal or opposite phases can be regarded as conjugation combined with the appropriate equivalence. After all, if $g_j = e^{i\theta} \tilde g_j$ with $\tilde g_j\in\RRR$ for all $j$, then $U^{-1}\psi$ evolves with $H_{\tilde g}$, which is time symmetric with time reversal given by conjugation. Put differently, $(H_g)^\rev=H_{g^*}$ as noted already in \eqref{HgHg*}, and $H_{g^*}$ is (in this case of equal or opposite phases) physically equivalent to $H_g$. So, one can say that up to equivalence, $T$ is conjugation, and that up to equivalence, time reversal conjugates the charges.

\section{Effective Potential}
\label{sec:effective}

The exchange of bosons constitutes an interaction between the fermions. If the fermions move slowly, and if the bosons are in the ground state, then the bosons tend to remain in the ground state even though the ground state changes slowly as it depends on the locations of the fermions. Moreover, the ground state energy as a function of the fermion coordinates acts as an effective potential for the motion of the fermions (see, e.g., \cite{T03}). We now compute this effective potential for $N$ complex charges.

The eigenvalue (i.e., ground state energy) of the ground state \eqref{groundstate} is
\be\label{Yukawa}
E_{\min}=\frac{m}{\pi\hbar^2}\biggl( \frac{\sqrt{2mE_0}}{2\hbar} \sum_{i=1}^N|g_i|^2-\sum_{1\leq i<j\leq N} \!\!\! \Re(g_i^*g_j)\frac{e^{-\sqrt{2mE_0}|\vx_i-\vx_j|/\hbar}}{|\vx_i-\vx_j|} \biggr)\,.
\ee
Regarding this energy function of $\vx_1,\ldots,\vx_N$ as an effective potential for the $x$-particles, we see that the $x$-particles effectively interact through Yukawa pair potentials,
\be
V(R)=\text{const.}-\kappa \frac{e^{-\lambda R}}{R}
\ee
with $R$ the distance between two $x$-particles, $1/\lambda$ the range of the interaction, and $\kappa$ the strength of the interaction. We find that $\lambda=\sqrt{2mE_0}/\hbar$, as originally obtained by Yukawa \cite{Yuk35} considering the effective interaction of nucleons by exchange of pions (except for a factor $\sqrt{2}$ presumably due to the non-relativistic nature of our model). We further find that
\be
\kappa_{ij}=\tfrac{m}{\pi\hbar^2} \Re(g_i^*g_j)\,,
\ee
so that, for fixed $|g_i|$ and $|g_j|$, the interaction strength is maximal when the complex charges $g_i$, $g_j$ have equal or opposite phases and vanishes for a phase difference of $\pm\pi/2$.

\section{Ordinary Boundary Conditions}
\label{sec:BC}

While IBCs generically lead to time asymmetry, we show in this section that ordinary boundary conditions, such as Dirichlet or Neumann conditions, do not. 

In contrast to IBCs, which relate values and derivatives of $\psi$ on the boundary to values of $\psi$ at interior points, ordinary boundary conditions involve only the values and derivatives of $\psi$ on the boundary. A \emph{local} boundary condition involves the value and derivatives of $\psi$ at \emph{only one boundary point}; a periodic boundary condition is an example of a non-local boundary condition. 

\subsection{Local Boundary Conditions}

As perhaps the simplest example of a space with boundaries, we consider the unit interval $\Q=[0,1]$; the associated Hilbert space $\Hilbert=L^2(\Q)$; and the Hamiltonian $H$ given by $-\tfrac{\hbar^2}{2m}\partial_x^2$ with general local boundary conditions
\begin{subequations}\label{BC1}
\begin{align}
(\alpha_0 + \beta_0 \partial_x)\psi(0) &= 0  \label{BC1a}\\
(\alpha_1+\beta_1\partial_x)\psi(1) &= 0 \label{BC1b}
\end{align}
\end{subequations}
with complex constants $\alpha_0,\alpha_1,\beta_0,\beta_1$ such that $(\alpha_0,\beta_0)\neq (0,0) \neq (\alpha_1,\beta_1)$. The Dirichlet condition is included in this scheme for $\beta_i=0$, the Neumann condition for $\alpha_i=0$.

Not every boundary condition makes $H$ self-adjoint and thus the time evolution unitary. Since unitarity is connected to the conservation of probability, it is not surprising that exactly those boundary conditions make $H$ self-adjoint that imply vanishing probability current into the boundary. The current into boundary point 1 is $j(1)$ with
\be
j(x) = \tfrac{\hbar}{m} \, \Im[\psi^*(x) \partial_x \psi(x)] ~.
\ee
If $\beta_1\neq 0$ then, by \eqref{BC1b},
\be
j(1) = \tfrac{\hbar}{m} \, \Im \bigl[\psi^*(1) (-\tfrac{\alpha_1}{\beta_1})\psi(1)\bigr]
= -\tfrac{\hbar}{m} |\psi(1)|^2 \, \Im \tfrac{\alpha_1}{\beta_1}\,.
\ee
To ensure conservation of probability, we need to choose $\alpha_1,\beta_1$ so that $j(1)=0$; so, we need that $\alpha_1/\beta_1\in\RRR$. If, however, $\beta_1=0$, then \eqref{BC1b} entails that $\psi(1)=0$ and automatically $j(1)=0$. That is, the boundary condition is either a Dirichlet condition
\be
\psi(1)=0
\ee
or of the form
\be\label{BC5}
\partial_x \psi(1) = \gamma_1 \psi
\ee
with real coefficient $\gamma_1$. Likewise, conservation of probabilities requires $j(0)=0$ and thus that the boundary condition is either of Dirichlet type,
\be
\psi(0)=0
\ee
or of the form
\be
\partial_x \psi(0) = \gamma_0 \psi
\ee
with real coefficient $\gamma_0$. It follows that both boundary conditions are invariant under complex conjugation, and thus that $H$ is time symmetric. The same happens in higher dimension: a Hamiltonian given by $-\Delta +V$ with real-valued potential $V$ and local boundary conditions is always time symmetric when it is self-adjoint.

\subsection{Point Interactions}

Points interactions (e.g., \cite{AGHKH88}) are, roughly speaking, Hamiltonians $H=-\Delta+V$ with potentials of the form $V(\vx)= g \,\delta^3(\vx-\vx_0)$, $g=g^*$. It is not surprising that they are time symmetric, given that $H=-\Delta+V$ with real-valued functions $V$ are. 

On a more precise level, point interactions in 3 dimensions are defined by the Bethe--Peierls boundary condition \cite{BP35},
\be\label{BP}
\lim_{r\to0+} (\partial_r - \gamma)\bigl( r \psi(r\vomega) \bigr)=0 
\ee
for all $\vomega\in\RRR^3$ with $|\vomega|=1$ (then $g=\eta+\gamma\eta^2/4\pi$ with infinitesimal $\eta$). Since the constant $\gamma$ is real, we see directly that  $\psi^*$ satisfies \eqref{BP} if $\psi$ does, and so we see, in very much the same way as for \eqref{BC5}, that point interactions are time symmetric.

\subsection{Non-Local Boundary Conditions}

Time symmetry properties are different for non-local boundary conditions, such as (still for $H\psi=-\tfrac{\hbar^2}{2m}\partial_x^2\psi$)
\begin{subequations}\label{BC2}
\begin{align}
\psi(1) &= e^{i\theta} \psi(0)   \label{BC2a}\\
\psi'(1) &= e^{i\theta} \psi'(0) \label{BC2b}
\end{align}
\end{subequations}
with $\theta\in\RRR$, which is a periodic boundary condition with a phase shift of $e^{i\theta}$. Self-adjointness and conservation of probability are related to the fact that \eqref{BC2} implies $j(1)=j(0)$, so any amount of probability lost at $x=1$ returns at $x=0$. (In Bohmian terms, a particle reaching $x=1$ jumps to $x=0$, while its velocity is continuous.) 

This model is equivalent to the following one which can be regarded as a simplified version of the Aharonov--Bohm effect \cite{AB59}, thus linking its time asymmetry (see below) to external magnetic fields, which, as mentioned, change sign under time reversal. Consider a circle of perimeter 1, let the wave function $\psi$ be a cross-section of a Hermitian rank-1 vector bundle $E$ over the circle, $\nabla_x$ the covariant derivative relative to a connection on $E$ that preserves the inner products in the fibers of $E$, and $H=-\tfrac{\hbar^2}{2m}\nabla_x^2$. The connection is uniquely determined, up to isomorphisms of $E$, by its holonomy, which must be a unitary endomorphism of a fiber space, that is, since fiber spaces are 1-dimensional, a complex number of modulus 1. This number can be identified with $e^{i\theta}$, and the equivalence with \eqref{BC2} arises by choosing an orthonormal basis (i.e., a unit vector) in some fiber and then transporting it around the circle using the parallel transport defined by the connection; the basis permits, at every point on the circle, an identification between the fiber space and $\CCC$.

The Hamiltonian with boundary conditions \eqref{BC2} is time \emph{asymmetric} whenever $\theta$ is not an integer multiple of $\pi$. Indeed, by \eqref{asym} the time reverse of $\psi$ must be $e^{i\theta(q)}\psi^*$ with locally constant (and thus, in this case, constant) $\theta(q)$. Since a global phase factor does not affect the Bohmian trajectories, it can be dropped, so there is no alternative to representing time reversal by $\psi\to\psi^*$. Since conjugation replaces $e^{i\theta}$ in \eqref{BC2} by $e^{-i\theta}$, time symmetry requires $e^{-i\theta}=e^{i\theta}$ or $\theta \in \pi\ZZZ$.

And again, we find a non-zero current in the ground state for a time asymmetric $H$. Indeed, the eigenfunctions are, for $-\pi<\theta<\pi$ with $\theta\neq 0$, $\psi=e^{ikx}$ with $k=\theta+2\pi n$, $n\in\ZZZ$, and (non-degenerate) eigenvalues $E=\hbar^2k^2/2m$. Thus, the ground state occurs for $k=\theta$, it is non-real and has non-zero current $j=\hbar k/m=\hbar\theta/m$; the Bohmian particle is moving along the circle at constant speed $\hbar\theta/m$.

\section{Conclusions}
\label{sec:conclusions}

We have shown for certain interior--boundary conditions (IBCs) that a generic choice of parameters leads to a violation of time reversal symmetry. While this means that such choices are unphysical for fundamental physical theories, they may well arise as effective models, and they are of interest precisely because time \emph{asymmetric} Hamiltonians are rather unfamiliar. We have discussed how time reversal needs to be represented in such theories, and have identified the time reversal operator $T$ by means of Bohmian trajectories also in cases in which $T$ is more than mere complex conjugation. While other kinds of arguments \cite{Uhl63,Rob} also yield information about $T$, the Bohmian theory allows for a particularly obvious, clear-cut, and direct approach to determining $T$. A Hamiltonian involving particle emission and absorption by sources with complex charges is time asymmetric iff not all charges have equal or opposite phases. We have also explored properties of the time asymmetric models, in particular the possibility of a non-vanishing current in the ground state.

\bigskip

\noindent\textit{Acknowledgments.} 
We are grateful to Stefan Keppeler for helpful discussions. J.S. received funding from the German Research Foundation (DFG) within the Research Training Group 1838 \textit{Spectral Theory and Dynamics of Quantum Systems}.


\begin{thebibliography}{29}

\bibitem{AB59} Y.~Aharonov and D.~Bohm:
	Significance of electromagnetic potentials in the 
	quantum theory. 
	{\em Physical Review (2)} {\bf 115}: 485--491 (1959) 

\bibitem{AGHKH88} S.~Albeverio, F.~Gesztesy, R.~H\o egh-Krohn, 
	and H.~Holden:
	\textit{Solvable models in quantum mechanics.}
	Berlin: Springer-Verlag (1988)

\bibitem{BM} M.V. Berry and R.J. Mondragon:
	Neutrino billiards: time-reversal symmetry-breaking 
	without magnetic fields.
	\textit{Proceedings of the Royal Society A} 
	\textbf{412}: 53--74 (1987)

\bibitem{BP35} H.~Bethe and R.~Peierls:
	Quantum Theory of the Diplon.
	\textit{Proceedings of the Royal Society of London A} 
	\textbf{148}: 146--156 (1935)

\bibitem{CT16} C.W.~Cowan and R.~Tumulka:
	Epistemology of Wave Function Collapse in Quantum Physics.
	\textit{British Journal for the Philosophy of Science} 
	{\bf 67(2)}: 405--434 (2016)
	\url{http://arxiv.org/abs/1307.0827}

\bibitem{bohmibc} D. D\"urr, S. Goldstein, S. Teufel, 
	R. Tumulka, and N. Zangh\`\i: 
	Bohmian Trajectories for Hamiltonians with 
	Interior-Boundary Conditions. 
	Perprint (2018)
	\url{http://arxiv.org/abs/1809.10235}

\bibitem{DGTZ04} D. D\"urr, S. Goldstein, R. Tumulka, and N. Zangh\`\i: 
	Bohmian Mechanics and Quantum Field Theory.
	\textit{Physical Review Letters} \textbf{93}: 090402 (2004)
	\url{http://arxiv.org/abs/quant-ph/0303156}

\bibitem{DGTZ05b} D. D\"urr, S. Goldstein, R. Tumulka, and N. Zangh\`\i: 
	Bell-Type Quantum Field Theories.
	\textit{Journal of Physics A: Mathematical and General} \textbf{38}: R1--R43 (2005)
	\url{http://arxiv.org/abs/quant-ph/0407116}

\bibitem{DT} D. D\"urr and S. Teufel:
	{\it Bohmian mechanics.} 
	Heidelberg: Springer-Verlag (2009)

\bibitem{KS15} S. Keppeler and M. Sieber: 
	Particle creation and annihilation at interior boundaries: one-dimensional models.
	{\it Journal of Physics A: Mathematical and Theoretical} {\bf 49}: 125204 (2016) 
	\url{http://arxiv.org/abs/1511.03071}

\bibitem{Lam18} J.~Lampart:
	A nonrelativistic quantum field theory with point 
	interactions in three dimensions. 
	Preprint (2018) 
	\url{http://arxiv.org/abs/1804.08295}

\bibitem{LS18} J. Lampart and J. Schmidt:
	On Nelson-type Hamiltonians and abstract boundary conditions. 
	To appear in \textit{Communications in Mathematical Physics} (2018)
	\url{http://arxiv.org/abs/1803.00872}

\bibitem{ibc2a} J. Lampart, J. Schmidt, S. Teufel, and R. Tumulka: 
	Particle Creation at a Point Source by Means of 
	Interior-Boundary Conditions. 
	{\it Mathematical Physics, Analysis, and Geometry} 
	forthcoming (2018)
	\url{http://arxiv.org/abs/1703.04476}

\bibitem{Mosh51a} M. Moshinsky:
	Boundary Conditions for the Description of Nuclear Reactions.
	{\it Physical Review} {\bf 81}: 347--352 (1951)

\bibitem{Mosh51b} M. Moshinsky:
	Boundary Conditions and Time-Dependent States.
	{\it Physical Review} {\bf 84}: 525--532 (1951)

\bibitem{Pen} R. Penrose: 
	\textit{The Emperor's New Mind.} 
	Oxford University Press (1989)

\bibitem{Rob} B. W. Roberts: 
	Three myths about time reversal in quantum theory.
	\textit{Philosophy of Science} {\bf 84}: 315--334 (2017)
	\url{http://arxiv.org/abs/1607.07388}

\bibitem{Sch18} J. Schmidt:
	On a Direct Description of Pseudorelativistic 
	Nelson Hamiltonians. 
	Preprint (2018)


\bibitem{T03} S. Teufel:
	\textit{Adiabatic perturbation theory in quantum dynamics.}
	Lecture Notes in Mathematics 1821. 
	Berlin: Springer-Verlag (2003) 

\bibitem{TT15a} S. Teufel and R. Tumulka:
	New Type of Hamiltonians Without Ultraviolet Divergence 
	for Quantum Field Theories.
	Preprint (2015)
	\url{http://arxiv.org/abs/1505.04847}

\bibitem{TT15b} S. Teufel and R. Tumulka:
	Avoiding Ultraviolet Divergence by Means of 
	Interior--Boundary Conditions.
	Pages 293--311 in F. Finster, J. Kleiner, C. R\"oken, 
	and J. Tolksdorf (editors), 
	\textit{Quantum Mathematical Physics -- A Bridge 
	between Mathematics and Physics.}
	Basel: Birkh\"auser (2016)
	\url{http://arxiv.org/abs/1506.00497}

\bibitem{Tho84} L.E. Thomas:
	Multiparticle Schr\"odinger Hamiltonians with 
	point interactions.
	{\it Physical Review D} \textbf{30}: 1233--1237 (1984)

\bibitem{Tum04} R. Tumulka and H.-O. Georgii:
	Some Jump Processes in Quantum Field Theory.
	    Pages 55--73 in J.-D. Deuschel and A. Greven (editors),
    \textit{Interacting Stochastic Systems}, Berlin:
    Springer-Verlag (2004). 
    \url{http://arxiv.org/abs/math.PR/0312326}

\bibitem{detect-rule} R. Tumulka:
	Distribution of the Time at Which an Ideal Detector Clicks.
	Preprint (2016)
	\url{http://arxiv.org/abs/1601.03715}

\bibitem{co1} R. Tumulka:
	Interior-Boundary Conditions for Schr\"odinger Operators on Codimension-1 Boundaries.
	Preprint (2018)
	\url{http://arxiv.org/abs/1808.06262}

\bibitem{Uhl63} U. Uhlhorn: 
	Representation of symmetry transformations in quantum 
	mechanics.
	\textit{Arkiv f\"or Fysik} \textbf{23}: 307--340 (1963)

\bibitem{Yaf92} D.R. Yafaev:
	On a zero-range interaction of a quantum particle with 
	the vacuum.
	{\it Journal of Physics A: Mathematical and General} 
	\textbf{25}: 963--978 (1992)

\bibitem{Yuk35} H. Yukawa:
	On the interaction of elementary particles. 
	\textit{Proceedings of the Physico-Mathematical Society 
	of Japan} \textbf{17}: 48--57 (1935) 

\end{thebibliography}
\end{document}